\documentclass[aps,prr,twocolumn,superscriptaddress]{revtex4-1}
\usepackage{graphicx}
\usepackage{epstopdf}
\usepackage{amsmath}
\usepackage{amssymb}
\usepackage{natbib}
\usepackage[latin1]{inputenc}
\usepackage{enumerate}
\usepackage{dashrule}
\usepackage{mathtools}
\usepackage{color,soulutf8}
\usepackage{tabularx}
\usepackage{longtable,array,url,textcomp}

\begin{document}

\title{Disorder-induced coupling of Weyl nodes in WTe$_2$}

\author{Steffen Sykora}
\affiliation{Leibniz Institute for Solid State and Materials Research (IFW Dresden), Helmholtzstraße 20, D-01069 Dresden, Germany}

\author{Johannes Schoop}
\affiliation{Leibniz Institute for Solid State and Materials Research (IFW Dresden), Helmholtzstraße 20, D-01069 Dresden, Germany}
\affiliation{Department of Physics, TU Dresden, D-01062 Dresden, Germany}

\author{Lukas Graf}
\affiliation{Leibniz Institute for Solid State and Materials Research (IFW Dresden), Helmholtzstraße 20, D-01069 Dresden, Germany}

\author{Grigory Shipunov}
\affiliation{Leibniz Institute for Solid State and Materials Research (IFW Dresden), Helmholtzstraße 20, D-01069 Dresden, Germany}

\author{Igor V. Morozov}
\affiliation{Leibniz Institute for Solid State and Materials Research (IFW Dresden), Helmholtzstraße 20, D-01069 Dresden, Germany}
\affiliation{Lomonosov Moscow State University, Moscow, 119991, Russia}

\author{Saicharan Aswartham}
\affiliation{Leibniz Institute for Solid State and Materials Research (IFW Dresden), Helmholtzstraße 20, D-01069 Dresden, Germany}

\author{Bernd B\"{u}chner}
\affiliation{Leibniz Institute for Solid State and Materials Research (IFW Dresden), Helmholtzstraße 20, D-01069 Dresden, Germany}
\affiliation{Department of Physics, TU Dresden, D-01062 Dresden, Germany}

\author{Christian Hess}
\affiliation{Leibniz Institute for Solid State and Materials Research (IFW Dresden), Helmholtzstraße 20, D-01069 Dresden, Germany}
\affiliation{Center for Transport and Devices, TU Dresden, D-01069 Dresden, Germany}

\author{Romain Giraud}
\affiliation{Leibniz Institute for Solid State and Materials Research (IFW Dresden), Helmholtzstraße 20, D-01069 Dresden, Germany}
\affiliation{Université Grenoble Alpes, CNRS, CEA, Spintec, F-38000 Grenoble, France}

\author{Joseph Dufouleur}
\affiliation{Leibniz Institute for Solid State and Materials Research (IFW Dresden), Helmholtzstraße 20, D-01069 Dresden, Germany}
\affiliation{Center for Transport and Devices, TU Dresden, D-01069 Dresden, Germany}
\date{\today}

\begin{abstract}
	The finite coupling between Weyl nodes due to residual disorder is investigated by magnetotransport studies in WTe$_2$. The anisotropic scattering of quasiparticles is evidenced from classical and quantum transport measurements. A new theoretical approach using the real band structure is developed in order to calculate the dependence of the scattering anisotropy with the correlation length of the disorder. For the first time, a comparison between theory and experiments reveals a short correlation length in WTe$_2$ ($\xi \sim 5$~nm). This result implies a significant coupling between Weyl nodes and other bands. Our study thus shows that a finite inter-cone scattering rate always exists in weakly-disordered type-II Weyl semimetals, such as WTe$_2$, which strongly suppresses topologically non-trivial properties.
\end{abstract}

\maketitle

\section{Introduction}

In Weyl and Dirac semimetals, bulk gapless excitations are described by a Dirac equation, with a linear band dispersion and a crossing at a band degeneracy point close to the Fermi energy. Contrary to Dirac semimetals, for which high symmetry Dirac points always give pairs of Fermions with opposite chiralities, the inversion symmetry breaking in Weyl semimetals splits the position of band degeneracy points in the reciprocal space into two distinct Weyl nodes. These Weyl nodes have opposite chiralities and are protected by the topology of the overall band structure. This gives rise to topologicaly non-trivial properties such as the presence of Fermi arcs \cite{Xu2015b,Xu2015a} or a chiral anomaly \cite{Son2013,Spivak2016,Gorbar2014}. Nevertheless, the observation of many topologically non-trivial properties requires a weak coupling between Weyl nodes or between the Weyl node and some other bands \cite{Son2013,Spivak2016}.

\begin{figure}[t]
	\includegraphics[width=0.9\columnwidth]{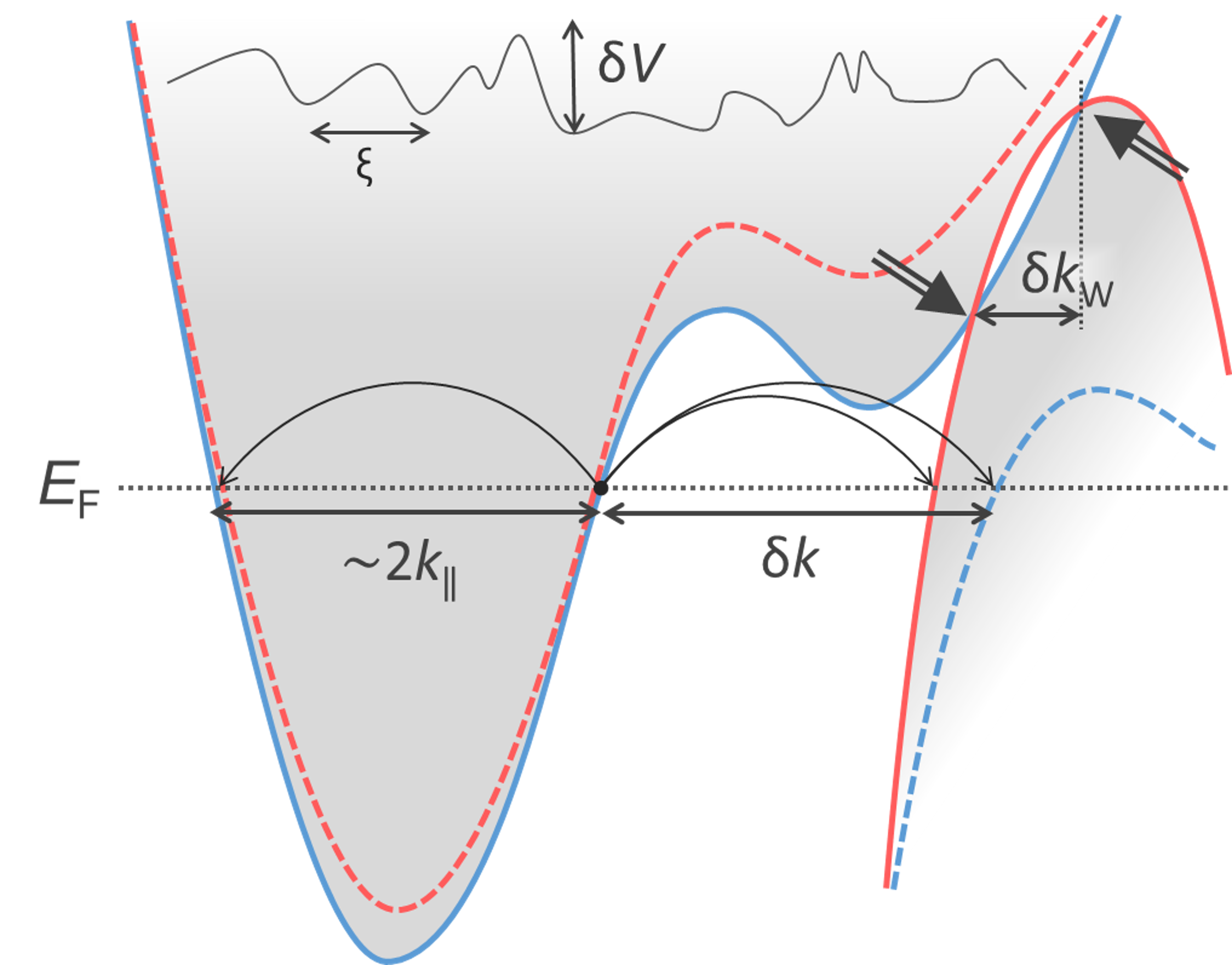}
	\caption[width=\textwidth]{Illustration of the band structure of WTe$_2$ with the Weyl nodes indicated by the two arrows. The spin degeneracies of the conduction and valence bands are lifted by the spin-orbit coupling, leading to two bands slightly shifted in energy (plain and dashed lines). The Fermi energy $E_\text{F}$ and the related effective Fermi wave vector $k_\parallel$ as defined in Eq.(\ref{Onsager}) are indicated for perfect compensation ($n=p$), giving rise to two spin non-degenerate electron and hole pockets contributing to the transport. An illustration of the disorder potential induced by impurities is given with a graphic definition of its amplitude $\delta V$ and its correlation length $\xi$.}
	\label{fig1}
\end{figure}

In the type-II Weyl semimetal WTe$_2$, the Weyl nodes are tilted such that the Fermi energy crosses the Weyl cones on both the electron and the hole sides \cite{Soluyanov2015} (see Fig.~\ref{fig1}). Besides the manifestation of a chiral anomaly \cite{Wang2016,Li2017}, this implies the presence of electron and hole pockets that touch each other when the Fermi energy is at the Weyl node. As electrons and holes are almost perfectly compensated in WTe$_2$, this results in the giant magnetoresistance reported in high quality  crystals \cite{Ali2014}. Recently, monolayers or few-layer thin films of WTe$_2$ attracted attention with the discovery of the quantum spin Hall effect  \cite{Qian2014,Peng2017,Tang2017,Fei2017,Jia2017,Xiang2018,Xu2018,Shi2019b}, the evidence of superconductivity \cite{Asaba2018,Fatemi2018,Sajadi2018} and the measurement of a non-linear Hall effect \cite{Kang2019,Ma2018}. Although the disorder in WTe$_2$ plays a key role in the observation of many trivial and non-trivial transport properties and despite some theoretical work \cite{Zhang2016b,Chen2016,Behrends2017,Lu2017,Ji2017,Knoll2019}, a quantitative and experimental study of the disorder in WTe$_2$ is still lacking. 

In this work, we investigate the magnetotransport properties of a WTe$_2$ nanoflake and, for the first time, we determine the correlation length $\xi$ of disorder, which determines the long- or short-range nature of the disorder as well as the coupling between the Weyl nodes. A comparison with a theoretical model that considers scattering processes within the specific band structure of WTe$_2$, using a renormalization method, allows us to infer a rather short $\xi$ ($\sim 5$~nm) of the disorder. This disorder leads to a finite coupling between Weyl nodes that can suppress the transport properties related to the topology of the band-structure.

In real materials, a particle can be scattered from a Weyl cone to another band due to the presence of impurities. This effect is related to the strength of the disorder ($\delta V$ in Fig.~\ref{fig1}) and to $\xi$ which defines the range of the scattering in reciprocal space. In a single-band model with a Fermi wave vector $k$, the short-range disorder limit is defined by $k \xi \ll 1$. Scattering is isotropic, \textit{i.e.} an initial state can be scattered all over the band and the angle $\theta_{\bf q}$ between initial (${\bf k}$) and final (${\bf k-q}$) states is uniformly distributed between 0 and 2$\pi$. On the other hand, in the long-range disorder limit corresponding to $k \xi \gg 1$, the disorder only couples states close to each other in reciprocal space and $\theta_{\bf q}$ is limited to small values (anisotropic scattering) \cite{DasSarma1985}. Similarly, in the multiband case, when considering the distance $\delta k$ between an initial state and some other
bands (see Fig.~\ref{fig1}), the inequality $\delta k \xi \ll 1$ sets the limit where interband scattering becomes significant. Particularly, for $\delta k_\text{W} \xi \ll 1$ with $\delta k_\text{W}$ being the distance between two Weyl nodes, disorder induces a finite relaxation rate between two Weyl cones of opposite chirality, preventing the measurement of a topologically non-trivial effect. Thus, the correlation length $\xi$ appears to be a key parameter for the observation of topologically non-trivial properties in Weyl (or Dirac) semimetals.

In order to measure $\xi$, we investigated the quantum transport properties of WTe$_2$ at very low temperature ($T \gtrsim 100$~mK) and under magnetic field ($B < 6$~T). In order to increase the signal-to-noise ratio of our measurements, we measure a thin flake of WTe2 with similar charge densities and mobilities than what is measured in macroscopic crystals from similar growths. A careful study of the Shubnikov-de Haas oscillations allow us to determine both the quantum lifetime $\tau'_{\text{Q}}$ of holes (the ' notation refer below to hole time scales and mobility) and an effective Fermi wave vector $k_\parallel$ of all charge pockets. Additionally, the transport times of electrons ($\tau_{\text{tr}}$) and holes ($\tau'_{\text{tr}}$) can be obtained from longitudinal and Hall magnetoresistances measurements. Far from the top of the valence bands, our calculations show that the ratio $\tau'_{\text{tr}}/\tau'_{\text{Q}}$ does not depend on the strength of the disorder and is a function of $k_\parallel$ and $\xi$ solely. Our theoretical approach based on a material specific band structure (including the spin texture) establishes the correspondence between $\tau'_{\text{tr}}/\tau'_{\text{Q}}$, $k_\parallel$ for holes and $\xi$ in a disordered material and, for the first time, gives a quantitative estimation of $\xi$.

\section{Two-band model and transport parameters}
\label{2BandModel}

\begin{figure}[t]
	\centering
	\includegraphics[width=1\columnwidth]{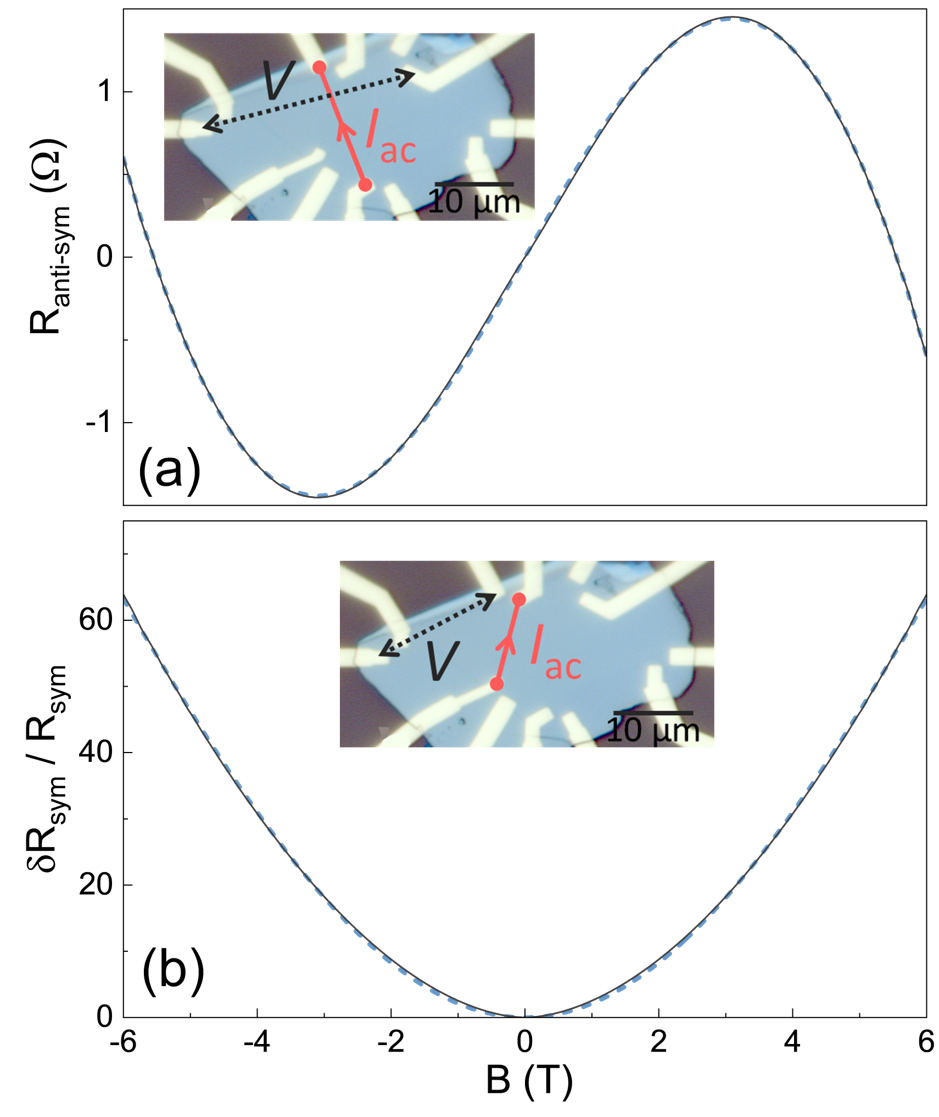}
	\caption[width=\columnwidth]{Anti-symmetrized magnetoresistance (a) of transverse contacts and symmetrized relative resistance (b) measured up to  $\pm 6$ T for longitudinal contacts. The black solid lines are the experimental data and the blue dashed lines are the fit to the non-compensated two-band model. The current and voltage contact configuration are indicated in the pictures.}
	\label{fig2}
\end{figure}

Bulk single crystals of WTe$_2$ were grown by Te flux and were characterized by SEM\textbackslash EDX mode for compositional analysis and with x-ray diffraction for structural analysis. A WTe$_2$ flake (about 30 $\mu$m $\times$ 50 $\mu$m $\times$ 70 nm) was directly exfoliated onto a Si/SiO$_x$ substrate and contacted by standard electron beam lithography and a metal lift-off process. Good ohmic contacts were obtained after in-situ ion beam etching prior to the electron beam evaporation of Cr(10 nm)/Au(100 nm) (see Fig.~\ref{fig2}).

When cooled down, the resistance decreases with a residual resistance ratio R$_\text{300K}$ / R$_\text{4K} \sim 80$, indicating a quality of our nanostructure comparable to what was reported so far in macroscopic crystals or in thin nanostructures {\cite{Ali2015,Na2016,Woods2017}}. At low temperature, a longitudinal magnetoresistance $\delta R/R=(R(B)-R(0))/R(0)$ exceeding 6000\% at $B=6$~T is measured (Fig.~\ref{fig2}). Hence, a tiny misalignement of Hall voltage probes results in a strong symmetric magnetoresistance which is not related to the anti-symmetric Hall voltage. In order to make a correct analysis, we therefore systematically symmetrized the longitudinal resistances and anti-symmetrized the transverse Hall resistances.

As shown in Fig.~\ref{fig2} both the longitudinal and the Hall resistivities can be described very well by the two-band model :
\begin{align}
\rho_\text{xx}&=\frac{1}{e}\frac{n\mu+p\mu'+(p\mu+n\mu')\mu\mu'B^2}{(n\mu+p\mu')^2+(p-n)^2\mu^2{\mu'}^2B^2}
\label{rho_xx}\\
\rho_\text{xy}&=\frac{B}{e}\frac{p{\mu'}^2-n\mu^2+(p-n)\mu^2{\mu'}^2B^2}{(n\mu+p\mu')^2+(p-n)^2\mu^2{\mu'}^2B^2}
\label{rho_xy}
\end{align}
with $n$ and $p$ being the respective charge densities of electrons and holes and $\mu$ and $\mu'$ their respective mobilities. The experimental results presented in Fig.~\ref{fig2} slightly depend on the contacts used which leads to a dispersion of about $\pm$~10\% in the measurments (see Appendix~\ref{ThetaDependence}). The Hall signal strongly deviates from the linear dependence that is expected in a fully compensated two-band model ($\delta n = n-p=0$). This is indeed expected from band calculations which predict the presence of four bands at full compensation, meaning two electron and two hole pockets. We note that the strong spin-orbit coupling lifts the spin-degeneracy of the different charge pockets that are nevertheless two times valley degenerated. Therefore, a departure from the fully compensated two-band model is anticipated by the theory. Still, the very good agreement with the non-compensated two-band model with $|\delta n| \ll n$ points to transport properties dominated by two effective bands with an almost perfect compensation between electrons and holes. In this approximation, the two spin non-degenerate electron (hole) pockets are considered as one electron (hole) single spin and valley-degenerated pocket.

The exact configuration of the current lines is not known in our nanostructure which prevents the use of geometrical parameters to fit the data with Eq.(\ref{rho_xx}). To overcome this difficulty, we fit $\delta R/R(B)=\delta \rho/\rho(B)$ instead of $R(B)$. Moreover, due to the almost complete compensation of charges ($|\delta n|/n \ll 1$), the term proportional to $\delta n^2$ in Eqs.(\ref{rho_xx}) and (\ref{rho_xy}) becomes too small to be reliably fitted. As a result, the fit of the magnetoresistances with Eqs.(\ref{rho_xx}) and (\ref{rho_xy}) allow us to determine three free parameters out of four unknown parameters ($n$, $p$, $\mu$ and $\mu'$). Therefore, the electron density is determined by the measurement of Shubnikov-de Haas oscillations as we will see below and we thus obtain $n= 3.01\times 10^{19}$~cm$^{-3}$. The fit of the magnetotransport data leads to $p=(3.05 \pm 0.3) \times 10^{19}$~cm$^{-3}$, $\mu=1.9 \pm 0.1$~m$^2/$(Vs) and $\mu'=1.1 \pm 0.1$~m$^2/$(Vs). We can also estimate the value of the transport time using the effective mass $m^*$ measured below: $\tau'_{\text{tr}}=m^*\mu'/e \simeq 2.5 \pm 0.3$~ps.

As a simple check from the transport parameters obtained, we can infer an effective electrical width $W$ that indicates the spread of the current lines between the source and the drain, and compare it to the typical size of the flake in the a-b plane. We have $W/L=[tR \times (e\mu_\text{tr} n + e\mu'_\text{tr}p)]^{-1}$ where $L$ is the source-drain distance, $t$ is the thickness of the flake and $R$ is the resistance at zero magnetic field. For $L=30 \mu$m, we obtain $W=29 \mu$m, a value that compares very well to the dimension of our flake.

\begin{figure}[t]
	\centering
	\includegraphics[width=1\columnwidth]{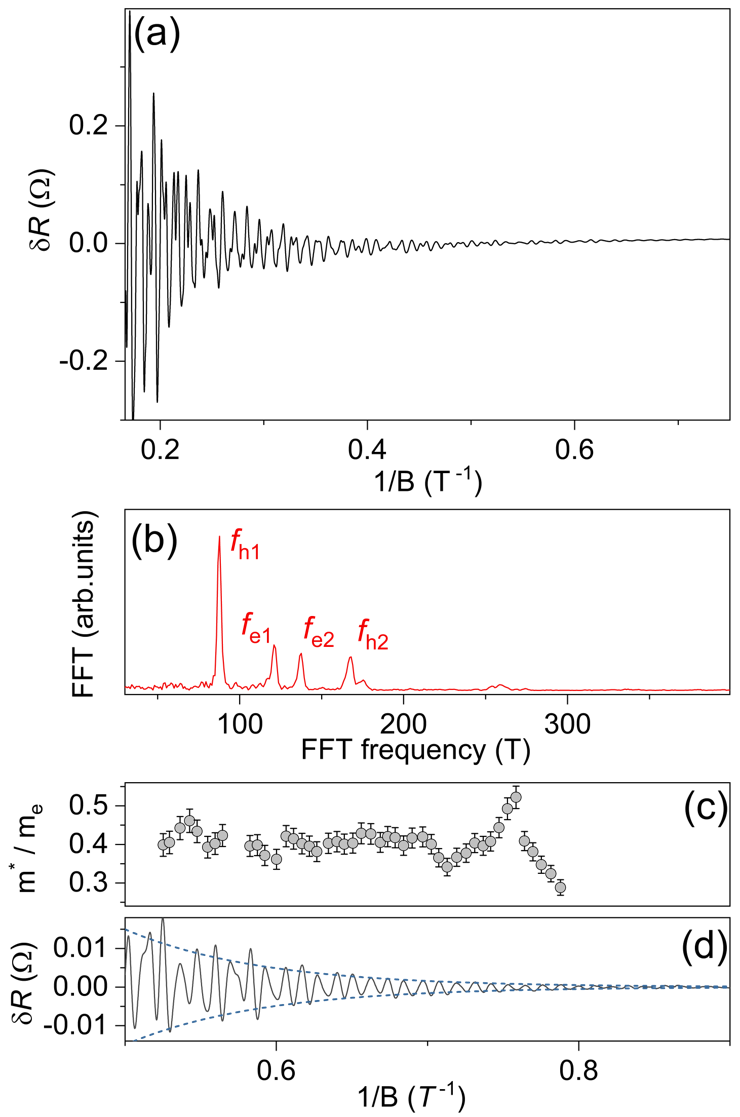}
	\caption[width=\columnwidth]{\textbf{a}: Shubnikov-de Haas oscillations without background measured in the longitudinal resistance at $T=100$~mK and up to 6~T. \textbf{b}: the amplitude of the Fourier coefficient of the quantum oscillations using a Blackman window. \textbf{c}:  The effective mass for the different extrema observed in $\delta R$ that leads to $m^* \simeq (0.40 \pm 0.03) \times m_\text{e}$. \textbf{d}: The dingle plot of the quantum oscillations  close to their onset. The blue dashed line indicates the best fit of the Dingle plot.}
	\label{fig3}
\end{figure}

\section{Quantum life time}

We focus now on the Shubnikov-de Haas oscillations measured down to $T \simeq$~100mK [Fig.~\ref{fig3}(a)]. The Fourier transform of the quantum oscillations reveals four sharp peaks (see fig.\ref{fig3}(b)) as expected from the band structure calculations \cite{Ali2014,Rhodes2015,Soluyanov2015} and as already observed experimentally \cite{Zhu2015,Rhodes2015}. It is straightforward to assign each peak to a given charge pocket by comparing our results with the theoretical expectations: the first ($f_\text{h1}=87$~T) and last peak ($f_\text{h2}=168$~T) correspond to the two hole pockets whereas the two other peaks at $f_\text{e1}=122$~T and $f_\text{e2}=135$~T are due to the contributions of the two electron pockets. Those values are in very good agreement with some previous work\cite{Zhu2015,Cai2015,Rhodes2015,Fatemi2017} as well as with our band structure calculation(see table~\ref{table}).

The frequencies of the quantum oscillations $f$ are related to the cross-section of the Fermi surfaces $A_\text{k}$ in the $k_\parallel$ plane by the Onsager relation,
\begin{equation}
	f=\frac{h}{4\pi^2 e} A_\text{k}=\frac{h}{4\pi e}{k_\parallel}^2
\label{Onsager}
\end{equation}
where $k_\parallel$ is the value of an effective in-plane Fermi wave vector assuming an isotropic in-plane band structure. In order to calculate the charge density associated to a given pocket, the value of the out-of-plane Fermi wave vector $ k_\perp $ is also required. Assuming an ellipsoidal pocket, the volume of the pocket is given by $ 4\pi k_\parallel^2 k_\perp /3$ and the associated charge density per spin by $ k_\parallel^2 k_\perp /6\pi^2 $. No Shubnikov-de Haas oscillations were measured for an in-plane magnetic field, making the determination of $k_\perp$ impossible. Therefore, we determine the value of $k_\perp$ for the different pockets from the band structure.
	
To do so, we first set the position of the Fermi energy such that the value of $A_\text{k}$ for the small hole pocket corresponds to the experimental value given by $f_\text{h1}$ in the fast-Fourier transform (FFT). We calculate then $A_\text{k}$ for the three other pockets (e1, e2 and h2) from the band structure and compare the calculated $k_\parallel$ with the experimental value given by $f_\text{e1}$, $f_\text{e2}$ and $f_\text{h2}$. The excellent agreement between experience and theory (see table~\ref{table} validates our band structure calculation and the exact position of the Fermi energy. We note that this position slightly differs from the one calculated for an undoped band structure, suggesting an intrinsic doping of the crystal. Finally, taking the value of $ k_\perp $, from the band structure calculation, we determine the charge density in a simple ellipsoidal approximation of the Fermi surfaces. Such an approximation is accurate within less than 1\% by comparison with a numerical approach that consists in counting the number of state present in the Fermi surfaces in the band structure. The electron density $n$ (from the FFT) is calculated by adding the contribution of the spin-degenerated $e1$ and $e2$ pockets, taking into account their valley degeneracy and we found $n=2n_\text{e1}+2n_\text{e2}=(3.01 \pm 0.03) \times 10^{19}$~cm$^{-3}$. We follow the same procedure to determine the hole density $p$ (from the FFT) of the  $h1$ and $h2$ pockets. We have $p=2p_\text{h1}+2p_\text{h2}=(3.00 \pm 0.03) \times 10^{19}$~cm$^{-3}$. The value of $p$ is consistent with the one obtained from the two-band model fit (fixing $n$ to the FFT value) that gives $p=(3.05 \pm 0.3) \times 10^{19}$~cm$^{-3}$ as already mentioned in the section~\ref{2BandModel}. As shown in table~\ref{table}, the almost perfect compensation confirms a posteriori the non-reliability of the term proportional to $\delta n^2$ in the fit of the two-band model ($ \delta n \ll n $) and the necessity of fixing the value of $n$ by independent measurement.

Close to their onset ($\simeq 1$~T), the quantum oscillations exhibit a single frequency $f$ in $1/B$ with $f=f_\text{h1}$, corresponding to a Fermi wave vector $k_\parallel \simeq 0.51 \pm 0.01$~nm$^{-1}$. These oscillations can clearly be assigned to the smallest hole pocket. The temperature dependence indicates a field independent effective mass $m^* \simeq (0.40 \pm 0.03) \times m_\text{e} $ where $m_\text{e}$ is the free electron mass [see appendix \ref{EffectiveMass} and  Fig.~\ref{fig3}(c)]. A fit of the extrema with a Dingle plot gives a quantum lifetime $\tau'_{\text{Q}} \sim 0.50 \pm 0.08$~ps [Fig.~\ref{fig3}(d)]. The contribution of different pockets at higher fields does not allow us to determine the effective mass or the quantum lifetime for the other pockets.

We determine the ratio $\tau'_{\text{tr}}/\tau'_{\text{Q}}$ for holes and find $\tau'_{\text{tr}}/\tau'_{\text{Q}} \simeq 5$. This ratio depends on the value of $k\xi$ and is therefore a measure for the anisotropy of the scattering. A large ratio stands for a large anisotropy and long-range disorder whereas a ratio close to one indicates an isotropic scattering and a short-range disorder. Theoretically, such a ratio can be much larger than one in topologically-trivial materials like two dimensional electron gases \cite{DasSarma1985}, graphene \cite{Hwang2008} or in topological insulators \cite{culcer2010} for instance. This emphasizes that this property is not related to any topological property of the band structure, but is rather only to the long-range nature of the disorder ($k\xi \gg 1$). Experimentally, a very large ratio (up to 60) was already measured in two dimensional electron gases \cite{Paalanen1983,Coleridge1989,Mancoff1996}. Such a large ratio is specific to the case of heterostructures where the disorder is located far from the metallic interface, therefore resulting in a very long-range disorder. In graphene, the ratio remains below two \cite{Monteverde2010} whereas it can reach eight in disordered 3D topological insulators \cite{Dufouleur2016}. In WTe$_2$, the values reported so far are between one and two for similar nanostructures \cite{Na2016,Woods2017}, significantly smaller than the value reported in the present work.

\begin{table*}
	\caption{FFT frequencies of the different peaks, experimental values of $k_\parallel$ [from Eq.(\ref{Onsager})], theoretical values of $k_\parallel$ (from the band structure, fixing the value of h1 to its experimental value), $k_\perp$ from the band calculation, densities calculated for the different charge pocket without any valley degeneracy, total charge density from the FFT (with valley degeneracy), total charge density from the two-band model ($n$ is fixed to its FFT value), transport mobility from the two-band model.}
	\begin{ruledtabular}
		\begin{tabular}{ccccccccc}
			&FFT & $k_\parallel$ & $k_\parallel$ & $k_\perp$ & $n$ & $n_\text{tot}$ & $n_\text{tot}$  & $\mu$\\
			& & exp. & theory & theory & per pocket & from FFT & two-band model &\\
			& T & (nm$^{-1}$) & (nm$^{-1}$) & (nm$^{-1}$) & $\times 10^{18}$ (cm$^{-3}$) & $\times 10^{18}$ (cm$^{-3}$) & $\times 10^{18}$ (cm$^{-3}$) & (m$^2$.V$^{-1}$.s$^{-1}$)\\
			\hline
			\rule{0pt}{2.5ex}
			h$_1$ or h & 87  & 0.51 & 0.51 (from exp.)& 1.08 & 4.82  & 30.0 & 30.5 & 1.14\\
			e$_1$ or e & 122 & 0.61 & 0.59 & 1.14 & 7.13  & 30.1 & 30.1 (from FFT)& 1.86\\
			e$_2$ & 135 & 0.64 & 0.66 & 1.14 & 7.89  & - & - & - \\
			h$_2$ & 168 & 0.72 & 0.71 & 1.18 & 10.17 & - & - & - \\
		\end{tabular}
	\end{ruledtabular}
	\label{table}
\end{table*}

\section{Theoretical model}

In order to extract the correlation length $\xi$ of the disorder, it is necessary to compare the experimental ratio $\tau'_{\text{tr}}/\tau'_{\text{Q}}$ with the corresponding value calculated from an appropriate theoretical model. We developed therefore the realistic theoretical approach based on a material's specific band structure, including the spin-texture, and a quantitative treatment of the disorder scattering. To achieve our aim, we applied the projective renormalization method (PRM) \cite{BHS_2002} usually used for many-particles system to a realistic Hamiltonian of WTe$_2$ with a static disorder. The method will be presented in more details in Ref.\onlinecite{Sykora2020} and we describe it briefly below.

We first consider the minimal case of WTe$_2$ in the presence of disorder $V$ where two bands interact with each other within the nodal energy crossings in the Brillouin zone. The corresponding Hamiltonian reads
\begin{eqnarray}
\label{H_scattering}
\mathcal{H} &=& \sum_{{\bf k},\alpha,\beta}  c_{{\bf k},\alpha}^\dagger \, [\hat{H}_{\bf k}]_{\alpha,\beta} \, c_{{\bf k},\beta}^{} \nonumber \\
&+& \sum_{{\bf k},{\bf k}',\alpha,\beta} \left( c_{{\bf k},\alpha}^\dagger [\hat{V}_{{\bf k},{\bf k}'}]_{\alpha,\beta} c_{{\bf k}',\beta}^{} + \mbox{h.c.} \right).
\end{eqnarray}
The operator $c_{{\bf k},\alpha}^\dagger$ creates an electron with momentum ${\bf k}$ and spin $\alpha$, i.e. $\alpha = \{\uparrow,\downarrow\}$ and the two-band Hamiltonian can easily be generalised to more bands by simply adding additional band indices. For a Weyl semimetal, there exists at least one pair of points in the Brillouin zone where the $2 \times 2$ matrix $\hat{H}_{\bf k}$ becomes linear in momentum. The scattering by the disorder is described by the $V$ term that relates an initial state $({\bf k}, \alpha)$ to a final state $({\bf k}', \beta)$. We take a Gaussian disorder\cite{Dufouleur2018} entirely characterized by the Fourier transform of its correlation function
\begin{eqnarray}
\label{init_V}
[\hat{V}_{{\bf k},{\bf k}'}]_{\alpha,\beta} = V e^{-2\xi^2|\mathbf{k}-\mathbf{k}'|^2} \delta_{\alpha,\beta},
\end{eqnarray}
where $V$ stands for the strength of the disorder and $\xi$ stands for its correlation length.

In our method, the non-diagonal Hamiltonian $\mathcal{H}$ is first decomposed into a diagonal part $\mathcal{H}_0$ and a non-diagonal part $\mathcal{H}_1$, i.e.~$\mathcal{H} = \mathcal{H}_0 + \mathcal{H}_1$. For this, we introduce new fermionic operators $a_{{\bf k},\alpha} = \sum_{\beta} [\hat{D}_{\bf k}]_{\alpha,\beta} c_{{\bf k},\beta}$ where the matrix $\hat{D}_{\bf k}$ is defined such that the first term in Eq.\eqref{H_scattering}  becomes diagonal: 
\begin{eqnarray}
\label{H_scattering_diag}
\mathcal{H} &=& \mathcal{H}_0 + \mathcal{H}_1 
= \sum_{{\bf k},\alpha} E_{{\bf k},\alpha} \, a_{{\bf k},\alpha}^\dagger a_{{\bf k},\alpha}^{} \nonumber \\
&+& \sum_{{\bf k},{\bf k}',\alpha,\beta} \left( a_{{\bf k},\alpha}^\dagger [\hat{D}_{\bf k} \hat{V}_{{\bf k},{\bf k}'} \hat{D}_{{\bf k}'}^{-1}]_{\alpha,\beta} a_{{\bf k}',\beta}^{} + \mbox{h.c.} \right).
\end{eqnarray}
All information about the spin texture band structure is now contained in the matrix $\hat{D}$. The decomposition \eqref{H_scattering_diag} of the Hamiltonian allows the application of the PRM to integrate out the scattering term using unitary transformations and to write $\mathcal{H}$ as the effective diagonal Hamiltonian $\tilde{\mathcal{H}}$ of free fermions with renormalized bands:
\begin{eqnarray}
\label{H_ren}
\tilde{\mathcal{H}} = e^{X} \mathcal{H} e^{-X} &=& \sum_{{\bf k},\alpha} \tilde{E}_{{\bf k},\alpha} \, \tilde{a}_{{\bf k},\alpha}^\dagger \tilde{a}_{{\bf k},\alpha}^{}  
\end{eqnarray}
with $X^\dag = -X$. Since $\tilde{\mathcal{H}}$ is connected to the original Hamiltonian $\mathcal{H}$ through a unitary transformation, it has the same eigenvalues as $\mathcal{H}$ \cite{SHB_2006_1} and allows access to any quantity of the system.

The parameters $\tilde{E}_{{\bf k},\alpha}$ are connected to the initial band structure $E_{{\bf k},\alpha}$ and initial scattering matrix elements $\hat{V}_{{\bf k},{\bf k}'}^a = \hat{D}_{\bf k} \hat{V}_{{\bf k},{\bf k}'} \hat{D}_{{\bf k}'}^{-1}$ through renormalization equations which are derived analytically within the scheme illustrated in Refs. \onlinecite{BHS_2002,SHBWF_2005}. The renormalization equations are solved numerically on a 3D grid of $20 \times 20 \times 20$ ${\bf k}$ points taking as starting values the realistic band structure of WTe$_2$ from Refs. \cite{McCormick2017,Liu2013} with a disorder described by a momentum distribution according to Eq.\eqref{init_V}.

The numerical solution of the renormalization equation allows to calculate the momentum dependent transport and quantum lifetimes given by Fermi's golden rule,
\begin{eqnarray}
\label{tau_Q}
\frac{1}{\tau_Q(\mathbf{k})} &=& \frac{2\pi}{\hbar} \sum_{\alpha,\beta,{\bf k}'} |\langle a_{\mathbf{k},\alpha}^\dagger [\hat{V}_{{\bf k},{\bf k}'}^a]_{\alpha,\beta}  a_{\mathbf{k}',\beta}^{} \rangle |^2 \nonumber \\
&\times & \delta(\tilde{E}_{{\bf k},\alpha} - \tilde{E}_{{\bf k}',\beta}), \\
\label{tau_tr}
\frac{1}{\tau_\text{tr}(\mathbf{k})} &=& \frac{2\pi}{\hbar} \sum_{\alpha,\beta,{\bf k}'} |\langle a_{\mathbf{k},\alpha}^\dagger [\hat{V}_{{\bf k},{\bf k}'}^a]_{\alpha,\beta}  a_{\mathbf{k}',\beta}^{} \rangle |^2 \nonumber \\
&\times & (1 - \cos \theta_{\bf q}) \delta(\tilde{E}_{{\bf k},\alpha} - \tilde{E}_{{\bf k}',\beta}),
\end{eqnarray}
with ${\bf q} = {\bf k} - {\bf k}'$ and $\theta_{\bf q}$ the angle between ${\bf k}$ and ${\bf k}'$. The scattering expectation values $\langle a_{\mathbf{k},\alpha}^\dagger  a_{\mathbf{k}',\beta}^{} \rangle = \langle \tilde{a}_{\mathbf{k},\alpha}^\dagger  \tilde{a}_{\mathbf{k}',\beta}^{} \rangle$ are calculated within the PRM using the same unitary transformation as for the renormalization of the Hamiltonian $\tilde{a}_{\mathbf{k},\alpha}^\dagger = e^{X} a_{\mathbf{k},\alpha}^\dagger e^{-X}$. 

\begin{figure}[t]
	\centering
	\includegraphics[width=1\columnwidth]{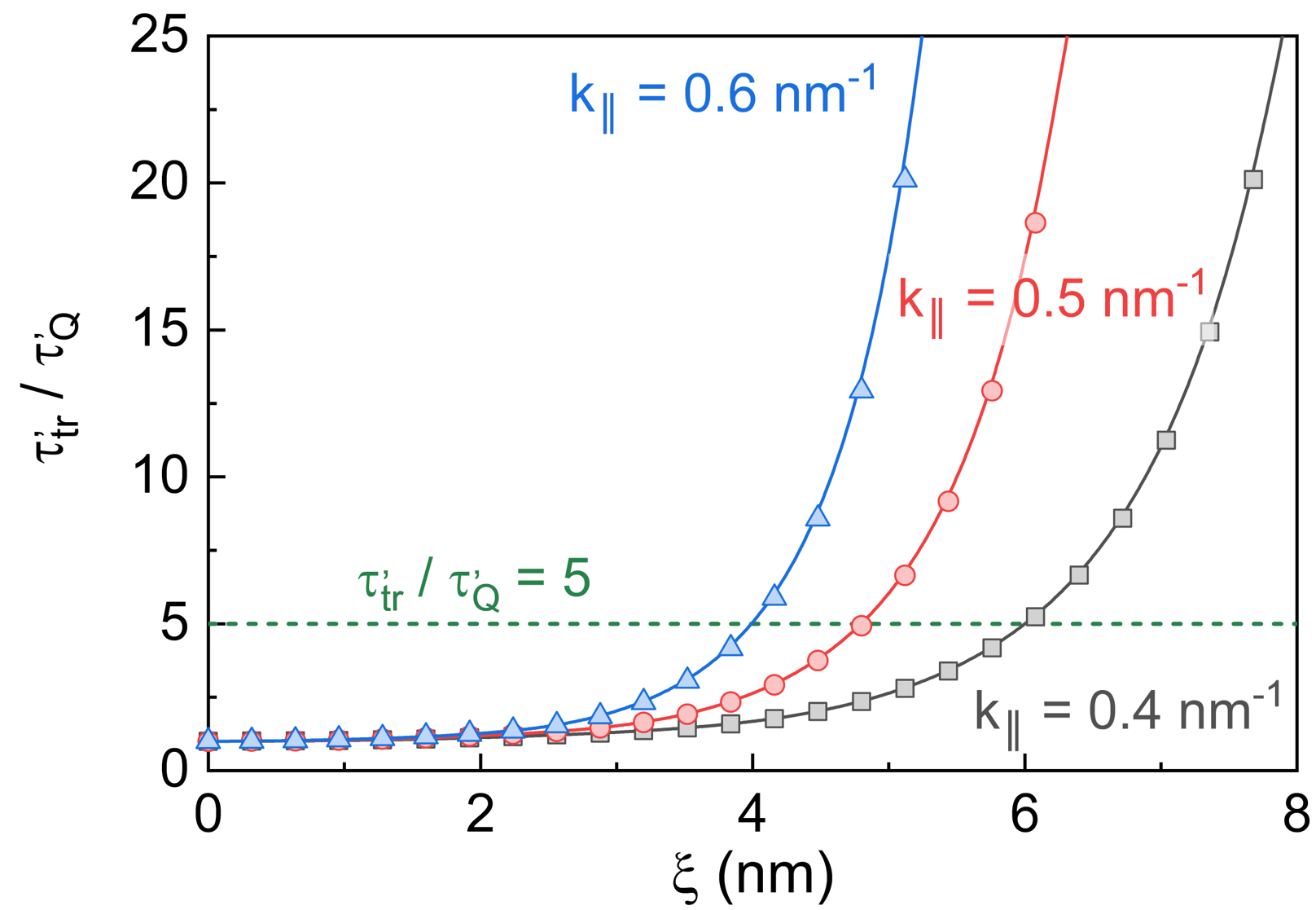}
	\caption[width=1\columnwidth]{Ratio $\tau'_\text{tr}/\tau'_\text{Q}$ calculated from the PRM where the value of $k_\parallel$ is fixed and the result is shown as a function of the correlation length $\xi$. The results for the  experimental values $k_\parallel = 0.5$~nm$^{-1}$ obtained from the quantum oscillation measurements is shown in red color together with the results corresponding to $k_\parallel = 0.4$~nm$^{-1}$ (black) and $k_\parallel = 0.6$~nm$^{-1}$ (blue). The dashed green line indicates the experimental value for $\tau'_\text{tr}/\tau'_\text{Q}$.}
	\label{fig4}
\end{figure}

Using the described approach we have evaluated numerically the expressions \eqref{tau_Q} and \eqref{tau_tr} for fixed ${\bf k} = k_\text{x}{\bf e}_x$ and $T=0$ where the value of $k_\text{x}$ corresponds to a given $k_\parallel$ as defined in Eq.(\ref{Onsager}). We present here the results obtained for ${\bf k}$ belonging to an hole Fermi pocket but our conclusions are not significantly changed if ${\bf k}$ belongs to a electron Fermi pocket. We also considered ${\bf k}$ to be directed in the other directions ${\bf e}_y$ and ${\bf e}_z$ but, again, the conclusions from numerical results were not affected. Finally, we chose for the strength of the disorder $V=$~5meV. Nevertheless, the ratio $\tau'_{\text{tr}}/\tau'_{\text{Q}}$ depends on $V$ only for small $k_\parallel$. Hence, the ratio calculated for $V=5$~meV and $V=50$~meV is the same as long as $k_\parallel \gtrsim 0.35$~nm$^{-1}$ and for the experimental value measured ($k_\parallel = 0.5$~nm$^{-1}$), the value of $V$ has no influence on our conclusions. 

In Fig.~\ref{fig4}, we plot the calculated ratio $\tau'_\text{tr} / \tau'_\text{Q}$ as a function of $\xi$ where $k_\parallel$ is defined as in Eq.(\ref{Onsager}). In the low $\xi$ limit, the short range disorder couples a state to all available states of the different Fermi pockets. Hence, the situation is almost equivalent to an isotropic scattering in spin degenerated bands and the ratio $\tau'_\text{tr} / \tau'_\text{Q} \sim 1$ as expected. In the opposite limit, for large $\xi$, the long-range disorder induces a strong anisotropic scattering and $\tau'_\text{tr} / \tau'_\text{Q}$ rises significantly. The limit between the two regimes is given by $k_\parallel\xi \sim 1$.

\section{Discussion and conclusion}

The comparison of the calculated ratio for $k_\parallel=0.5$~nm$^{-1}$ with the experimental value we measured gives $\xi \simeq 5$~nm. The value of $k_\parallel \xi \simeq 2.5$ rules out a disorder dominated by point-like impurities, for which $k_\parallel \xi \rightarrow 0$, and suggests a long-range disorder dominated by screened charged impurities \cite{culcer2010}. We stress that the dependence of the measurements on the contacts geometry can lead to an error in the determination of the ratio $\tau'_\text{tr} / \tau'_\text{Q}$ of about $\pm 5\%$. Nevertheless, the weak dependence of $\xi$ on this ratio for $\tau'_\text{tr} / \tau'_\text{Q} \simeq 5$ implies that such an error only weakly changes the final value of $\xi$ (a few percent, see appendix \ref{ErrorXi}).

We note here that $\xi$ is measured at a Fermi energy which is about 50~meV below the Weyl nodes.
Moving the position of the Fermi energy changes both the size of electron and hole pockets in an opposite way such that the total charge density (electron and hole) should not change significantly at the Weyl nodes. Therefore, any change in the position of the Fermi level should not substantially modify the screening properties of charged impurities by the Fermi seas, so that the energy dependence of both the amplitude of the potential and the correlation length of the disorder is expected to be rather weak.

A relevant value of the distance between a Weyl node and some other band in reciprocal space is given by  $\delta k_\text{W}$. This quantity has been calculated in some previous works \cite{Chang2016,Soluyanov2015,Wang2016d,Li2017} and values were found between $0$ and $0.32$~nm$^{-1}$. According to the value of $\xi$ we measured, we have $\delta k \xi \sim 1$, meaning a substantial coupling between a Weyl node and some other band that strongly reduces the signature of topological properties. A larger value of $\delta k \xi$ in WTe$_2$ would exponentially enhanced the amplitude of the measured chiral anomaly \cite{Wang2016,Li2017}, a property only limited by the intervalley scattering time.

Qualitatively, the value we extract for $\xi$ might not be restricted to the case of WTe$_2$ but could be generalized to any topological material with similar charge density and disorder. It highlights the necessity to enhance the crystal quality or to increase the distance in the reciprocal space between Weyl nodes to measure topologically non-trivial properties. This might be achieved in WTe$_2$ by substituing the Tungsten atoms by Molybdenum atoms in the crystal structure, giving W$_{1-\text{x}}$Mo$_\text{x}$Te$_2$ \cite{Chang2016}.

In conclusion, we study the transport properties of an exfoliated flake of WTe$_{2}$ and measure the anisotropy of the scattering of electrons. We develop a new theoretical method taking into account the real band structure of WTe$_2$ with the aim of comparing our experimental data with the numerical simulations and we determine for the first time the correlation length of the disorder $\xi \sim 5$~nm. This value points to a significant coupling between the Weyl nodes and other bands, leading to a strong reduction of topological properties. Our results stress the importance of band structure engineering of Weyl semimetals to enhance the distance $\delta k_\text{W}$ between the Weyl nodes in order to investigate topological properties.

\begin{acknowledgments}
S.A. acknowledges financial support by the Deutsche Forschungsgemeinschaft (DFG) through the grant AS 523/4-1. J.D. acknowledges financial support by the Deutsche Forschungsgemeinschaft (DFG) through the SPP 1666 Topological Insulators program (Project DU 1376/2-2) and the Würzburg-Dresden Cluster of Excellence on Complexity and Topology in Quantum Matter - ct.qmat (EXC 2147, project-id 0392019). This project has received funding from the European Research Council (ERC) under the European Unions Horizon 2020 research and innovation program (grant agreement No 647276-MARS-ERC-2014-CoG). IVM, SA and BB thank DFG and RSF for financial support in the frame of the joint DFG-RSF project \textquoteleft Weyl and Dirac semimetals and beyond-prediction, synthesis and characterization
of new semimetals\textquoteright (project number: 405940956). We acknowledge financial support from the DFG through SFB 1143 (project-id 247310070). We would like to thank Teresa Tschirner and Arthur Veyrat for their careful reading of the manuscript.
\end{acknowledgments}

\appendix

\section{Contact dependence of the transport properties}
\label{ThetaDependence}

\begin{figure}[h]
	\centering
	\includegraphics[width=1\columnwidth]{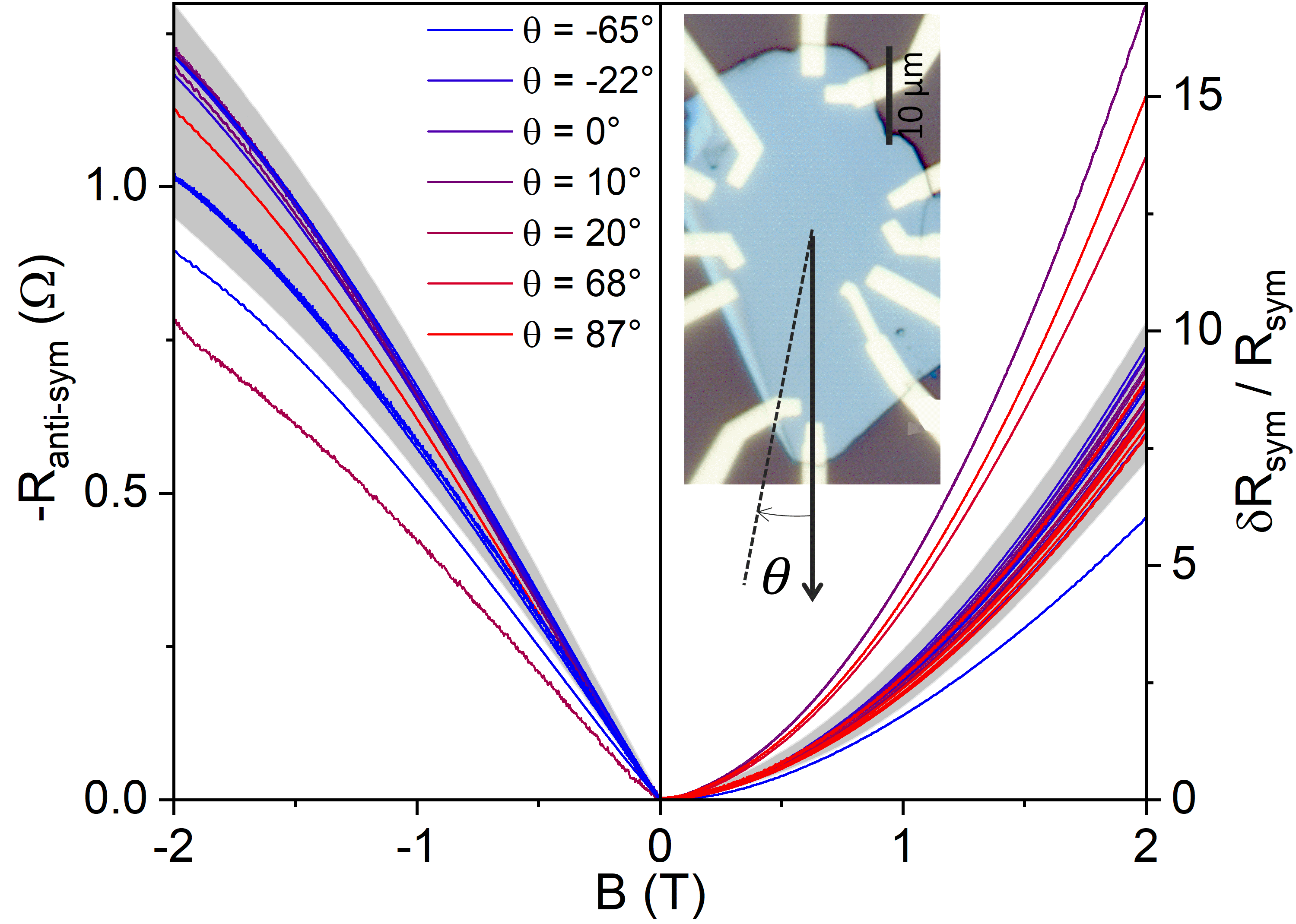}
	\caption[width=1\columnwidth]{Anti-symmetrized magnetoresistance and symmetrized relative resistance up to $\pm 2$ T for different orientations $\theta$ of the current and for different voltage probes and at different temperature between $T=4$ K and $T=100$ mK (no significant temperature dependence could be measured at low temperature). The $\theta$ values range between $-65^\circ$ and $87^\circ$ which is indicated by the change from blue to red. The grey zone roughly indicates where at least 80\% of the measurements can be found with a dispersion of about $\pm$ 10\%. A picture of the sample is shown in the inset of the right panel.}
	\label{FigThetaDependence}
\end{figure}

In order to measure the contact dependence of the magnetoresistance and the anisotropy, we measured the longitudinal and Hall magnetoresistance in different configurations at a magnetic field up to 2T. We measured 25 different configurations for the longitudinal magnetoresistance and 10 configurations for the Hall magnetoresistance in total. The results are shown in the Fig.\ref{FigThetaDependence} where the color of the different traces corresponds to the main orientation of the current lines (the $\theta$ angle between the reference axis in Fig.\ref{FigThetaDependence} and the source-drain axis). The discrepancies in the measurements are responsible for a dispersion of the experimental results of about $\pm$~10\%. Nevertheless, no systematic $\theta$ dependence could be evidenced so that no anisotropy could be revealed.

\section{Determination of the effective mass}
\label{EffectiveMass}

\begin{figure}
	\centering
	\includegraphics[width=1\columnwidth]{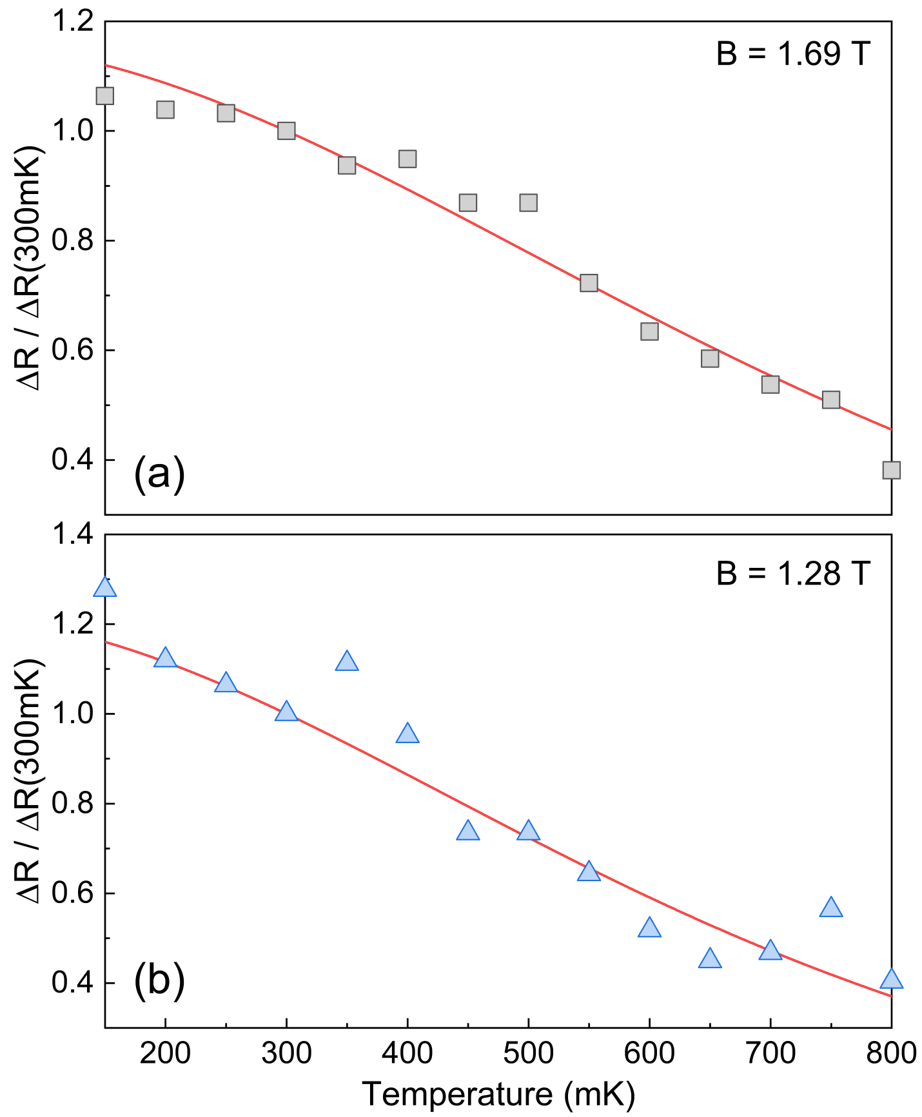}
	\caption[width=1\columnwidth]{Determination of the effective mass $m^*$ with the Lifshitz-Kosevich formulae at $B_\text{n}=1.69$~T (up) and $B_\text{n}=1.28$~T. The points are the values of the extrema and the red lines are the fits with the theoretical formulae.}
	\label{FigEffectiveMass}
\end{figure}

In order to determine the effective mass, we used the Lifshitz-Kosevich formulae \cite{Shoenberg2009}
\begin{equation}
\tag{A1}
\frac{\Delta R_\text{n}(T)}{\Delta R_\text{n}(T_0)}=\frac{T}{T_0}\frac{\sinh{\left[\left(2\pi^2 k_\text{B} m^* T_0\right)/\left(\hbar eB_\text{n}\right)\right]}}{\sinh{\left[\left(2\pi^2 k_\text{B} m^* T\right)/\left(\hbar eB_\text{n}\right)\right]}}
\label{Lifshitz}
\end{equation}
where we chose $T_0=300$~mK and $B_\text{n}$ to be the position of the different extrema in magnetic field. We restrict the study to a magnetic field where only a single frequency was clearly observed (1T<$B$<2T) and we show in Fig.\ref{FigEffectiveMass} two examples of such a thermal dependence close to the onset of the Shubnikov-de Haas oscillations ($B_\text{n}=1.28$~T) and at a higher field ($B_\text{n}=1.69$~T). The amplitude of the Shubnikov-de Haas oscillations being typically of about 1\textperthousand~of the signal at very low temperature, the noise in the measurement is relatively high which leads to a relatively large error bar but the effective mass is found to be $B$-independent [see Fig.\ref{fig3}(c)]. In order to reduce this error bar, we average $m^*$ over all the extrema. This results in $m^* \simeq (0.40 \pm 0.03) \times m_\text{e} $ where $m_\text{e}$ is the free electron mass.

\section{Error bar in the determination of $\xi$}
\label{ErrorXi}

As mentioned above, any error $\Delta r$ in the measurement of the ratio $r=\tau'_\text{tr}/\tau'_\text{Q}$ induces an error $\Delta \xi$ in the determination of $\xi$. As long as the errors remain reasonable, the ratio between the two relative errors is given by
\begin{equation}
\tag{A2}
\frac{\Delta \xi / \xi}{\Delta r / r}=\frac{d \xi}{d r}\frac{r}{\xi}
\end{equation}
This ratio of the two errors can be calculated from the data presented in Fig.\ref{fig4} and we plot the result as a function of the $\tau'_\text{tr}/\tau'_\text{Q}$ ratio. Interestingly, we note that the result is a function of $r$ but it does not depend on $k_\parallel$. Most importantly, one sees that for $r > 1.5$, the relative error induced in $\xi$ will be smaller than the one of $r$. Close to $r=5$, the experimental value we measure, the relative error in $\xi$ is even four times smaller than the one in $r$ which significantly enhances the accuracy of our measurement of $\xi$.

\begin{figure}[b]
	\centering
	\includegraphics[width=1\columnwidth]{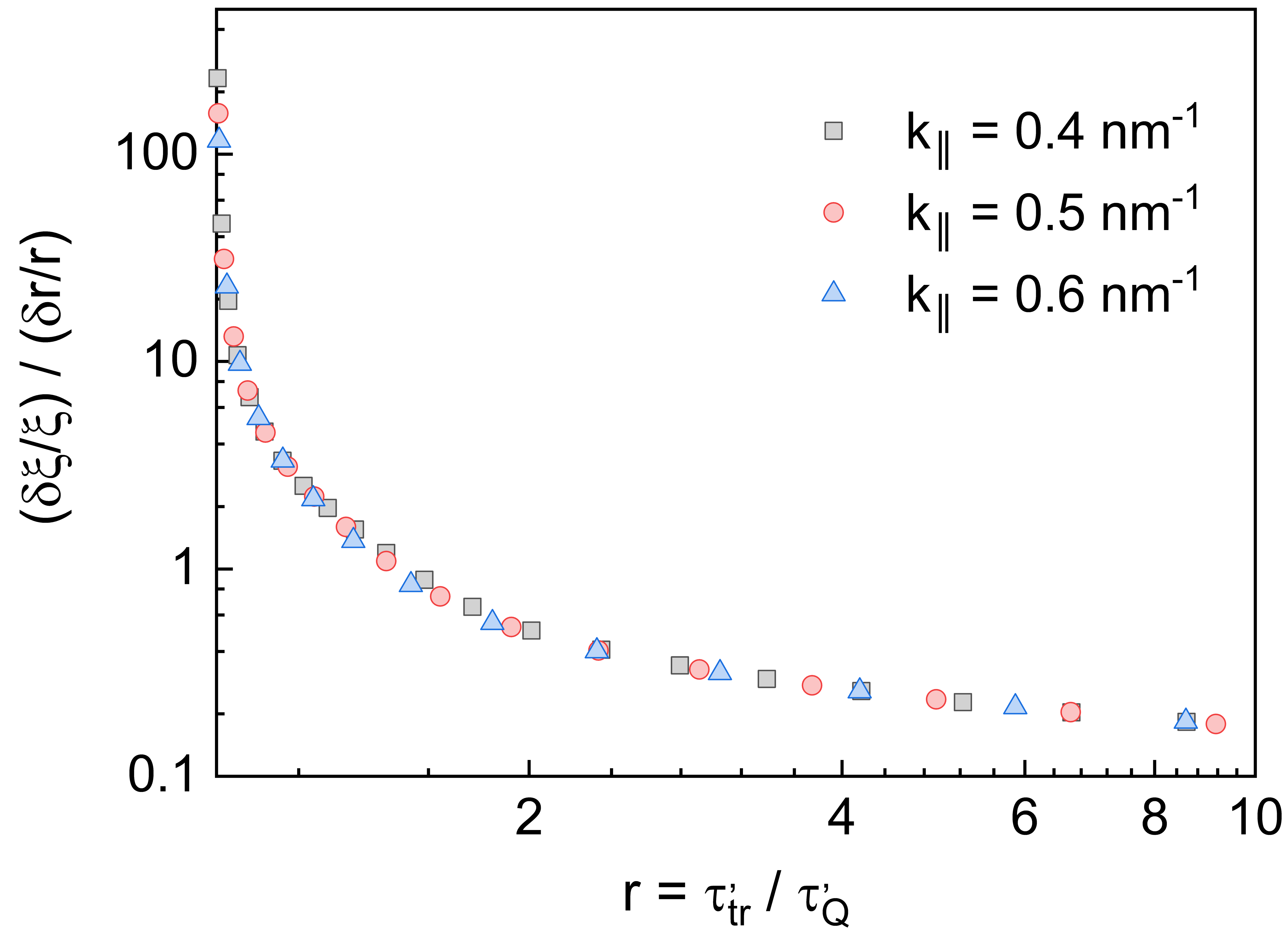}
	\caption[width=1\columnwidth]{Plot of the ratio of relative errors in $\xi$ and $r$ as a function of $r$ in a logarithmic scale. No $k_\parallel$ dependence could be observed.}
	\label{FigErrorXi}
\end{figure}

\bibliographystyle{apsrev4-1}
%

\end{document}